# Predictability of climate tipping focusing on the internal variability of the Earth system

Amane Kubo[1] & Yohei Sawada[1]

1: Department of Civil Engineering, The University of Tokyo

Corresponding to: Amane Kubo  amanetf0323@g.ecc.u-tokyo.ac.jp

**Abstract**

Prediction of climate tipping is challenging due to the lack of recent observation of actual climate tipping. Despite many previous efforts to accurately predict the existence and timing of climate tippings under specific climate scenarios, the predictability of climate tipping, the necessary conditions under which climate tipping can be predicted, has yet to be explored. In this study, the predictability of climate tipping is analyzed by Observation System Simulation Experiment (OSSE), in which the value of observation for prediction is assessed through the idealized experiment of data assimilation, using a simplified dynamic vegetation model and an Atlantic Meridional Overturning Circulation (AMOC) two box model. We find that the ratio of internal variability to observation error, or signal-to-noise ratio, should be large enough to accurately predict climate tippings. When observation can accurately resolve the internal variability of the system, assimilating these observations into process-based models can effectively improve the skill of predicting climate tippings. Our quantitative estimation of required observation accuracy to predict climate tipping implies that the existing observation network is not always sufficient to accurately project climate tipping.

**Key points**

- We discuss the practical predictability of climate tipping for the first time.
- Observation without any direct records of climate tipping can improve process-model based prediction of climate tipping if it is accurate enough.
- The skill of models to simulate internal variability leads to a successful prediction of climate tipping.

**Plain language summary**

Climate tipping is a critical and irreversible change in the earth system due to recent global warming, which is feared to have a huge impact on the society. To reduce the effect of climate tipping on society, it is important to know whether it will happen or not and to predict as early as possible if it will happen. However, the last climate tipping has occurred before the modern era and we do not have accurate records of climate tipping, which makes it difficult to predict it. Although many previous works used computer simulations to predict the existence and timing of climate tipping, it has yet to be clarified how observation data can contribute to this simulation-based assessment of climate tipping. Here we examine the predictability of climate tipping by identifying what kinds of observation data are necessary to improve the computer simulation of climate tipping. The ratio of year-to-year variability to observation error should be large enough to accurately predict climate tipping. Our results implies that the existing observation network is not always sufficient to accurately project climate tipping.

## 1. Introduction

Climate tipping is a sudden, critical and irreversible change of the earth system caused by recent global warming. For example, Amazon rainforest, which is the largest terrestrial carbon sink in the Earth system, is predicted to die off because it loses resilience against extreme climate under warmer climate (Cox et al.(2004)). Similar sudden changes of the Earth system have been pointed out, such as the Atlantic Meridional Overturning Circulation (AMOC) (Stommel (1961), Cessi (1994)), permafrost (Devoie et al. (2019)), and monsoon systems (Stolbova (2016)). It is a grand challenge to accurately predict the existence and timing of climate tippings under the current emission and warming scenarios.

There are two approaches to predict climate tipping: modeling and observation. In the modeling approach, many mathematical and numerical models have been developed. At the early stage, simple and stylized mathematical models were proposed, which successfully indicated tipping-like behaviors (Stommel (1961), North (1990)). These stylized models have been widely used to investigate the features of climate tipping (e.g., Ritchie et al. (2021), Lohmann et al. (2021), Wunderling et al. (2021)). Recently, Earth System Models (ESMs) such as CESM (Wasten and Djikstra (2023), Jackson and Wood (2018)) can also realistically reproduce the tipping-like behaviors. Observation networks to capture the signals of climate tipping have been growing (see McCarthy et al.(2020) for the comprehensive review of the AMOC strength observations). Scheffer (2009) proposed a statistical method to detect early signals of climate tipping (see also Clark et al.(2023), Westen et al.(2024), Boers (2021))

Despite these efforts to predict climate tippings, few studies have examined their predictability. Climate projection works so far focused on examining intrinsic predictability. Specifically, they investigated how small perturbations of the initial condition affect the outcome of climate simulation assuming that their model could perfectly reproduce the relevant processes of climate (e.g., Msadek et al. (2010), Weidman & Kuang (2023)). In the literature of climate tipping, previous works focused only on the intrinsic predictability using toy model (Ritche et al. (2021)). However, in practice, it is important to quantify how accurately the targeted phenomena can be predicted by imperfect models and how observation data can mitigate the impact of the imperfectness of models on prediction skills, which is related to practical predictability. What kinds of observation are necessary to constrain the trajectory of modelled Earth system accurately enough to predict climate tipping? In this paper, we answer this scientific questions by quantifying practical predictability of climate tipping.

In numerical weather prediction, practical predictability has been assessed by Observation System

Simulation Experiments (e.g., Zhang et al.(2016), Liu et al. (2018)). Practical predictability studies also have been done through simple model experiments to confirm the effectiveness of the new prediction method (Sawada (2022), Miwa and Sawada (2024)). In this paper, we examine practical predictability of climate tipping through its simple model experiments to find necessary condition of the system and observations to predict it.

2. Method

2.1 Observation System Simulation Experiment

Observation System Simulation Experiment (OSSE) is an experimental framework to investigate how much observations can improve model-based prediction through data assimilation. In OSSE, we virtually define a model as a 'true model' and obtain 'true state' called nature run by running the 'true model'. Then, we generate observation data from the nature run assuming some observation errors and imperfectness. We also define an 'imperfect model' which lacks some information about the 'true model'. We assimilate the observation data into the 'imperfect model' and evaluate how close the simulation by data assimilation is to the nature run. OSSE enables to validate the performance of data assimilation-based prediction more precisely than the real-world application, since we can directly compare data assimilation-based prediction with nature run, which is not available in reality. Please refer to Arnold Jr and Dey (1986) and Hoffman and Atlas (2016) for the complete description of OSSE.

2.2 Particle filter

Particle Filter (PF) is one of the data assimilation methods which estimate posterior distribution of model parameters and state variables when observations are given. PF has been widely used in geoscience (e.g., Moradkhani et al. (2005), Moradkhani et al. (2012), Poterjoy et al.(2019), Sawada et al. (2015), Sawada and Hanazaki (2020)). A discrete state-space model is defined as follows:

$$x(t) = f(x(t-1), \theta(t-1)) + \epsilon(t) \quad (1)$$

where $x(t)$ is the state variables, $\theta(t)$ is the model parameters and $\epsilon(t)$ is the noise process which represents the model error at time t. Equation (1) shows the dynamics of variable $x(t)$ depending on $\theta(t)$, which is defined as $f(x, \theta)$. Observation equation is defined as:

$$y(t) = h(x(t)) + \xi(t) \quad (2)$$

where $y(t)$ is observation and $\xi(t)$ is the noise process which represents the observation error at time t. Equation (2) shows the relationship between state variables and observation defined as $h(x)$. Based on the Bayesian update, posterior distribution of state variables and model parameters can be

derived by:

$$p(x(t), \theta(t)|y(1:t)) \propto p(y(t)|x(t), \theta(t)) \, p(x(t), \theta(t)|y(1:t-1)) \tag{3}$$

where $p(x(t), \theta(t)|y(1:t))$ is the posterior distribution of the state variables and model parameters given all observations up to time t, $y(1:t)$. The prior distribution $p(x(t), \theta(t)|y(1:t-1))$ is updated by the likelihood, $p(y(t)|x(t), \theta(t))$. PF approximates the prior distribution using Monte-Carlo methods. The complete description and hyperparameters of our PF can be found in the Supporting Information Text S1.

## 2.3 Climate tipping models
### 2.3.1 Simplified TRIFFID model

Top-down Representation of Interactive Foliage and Flora Including Dynamics (TRIFFID) is a global dynamic vegetation model (Cox et al. (2002)). Ritche et al. (2021) simplified this model to describe the tipping of the Amazon rainforest. The model's equations are as follows:

$$\frac{dv}{dt} = gv(1-v) - \gamma v, \tag{4}$$

$$g = g_0 \left[ 1 - \left( \frac{T_l - T_{opt}}{\beta} \right)^2 \right], \tag{5}$$

$$T_l = T + \alpha(1-v), \tag{6}$$

where $v$ is a tropical forest ratio to all the vegetation in Amazon, $T_l$ is temperature in the Amazon forest and $g$ is a growth ratio of rainforest. Equation (4) describes the dynamics of $v$ based on the Lotka-Volterra equation. The first term on the righthand side of Equation (4) describes the growth velocity of tropical forest ratio and the second term means the decrease velocity of the ratio where $\gamma$ is the disturbance ratio. Equation (5) shows how the growth ratio is determined depending on the temperature in the forest. The growth ratio reaches its maximum at the optimal temperature $T_{opt}(°C)$ and a reference temperature of the growth-ratio decrease is $\beta$ (°C). $T(°C)$ in Equation (6) is temperature at forest-covered areas in Amazon. $T(°C)$ is a forcing of this model. Equation (6) means the more dominant rainforest is in the Amazon forest, the lower the temperature in the Amazon forest is. $\alpha$ (°C) is a reference temperature of coverage-induced temperature change. The setting of the parameters of this model in this paper can be found in Table S2 and Table S3 in Supporting information.

We used the 4[th]-order Runge-Kutta method for numerical calculation. Timestep $dt$ was set to 0.1 year and the simulation period is 1000 years. Whether the system tips or not was determined by the criteria

$v < 0.1$ at the last step of the simulation.

### 2.3.2 AMOC two box model

The AMOC two box model describes the AMOC dynamics using two boxes which have different temperature and salinity (Stommel (1961), Cessi (1994)). The model's equations are the following:

$$\frac{dy}{dt} = F - y[1 + \mu^2(1-y)^2], \tag{7}$$

$$Q = \frac{\eta V[1 + \mu^2(1-y)^2]}{t_d}, \tag{8}$$

where $y$ is the difference of salinity between two boxes, Q (Sv) is the AMOC strength. Equation (7) describes the dynamics of salinity flux. F is external salinity flux, and the second term of the right-hand side of Equation (7) shows how much salinity difference and temperature difference between two boxes influence the change of AMOC strength in total. $\mu$ is the parameter of the ratio of the diffusion timescale $t_d$(year) to the advection timescale, $V(m^3)$ is the parameter of seawater volume, and $\eta$ is a scaling parameter of this model. In this model, the external forcing is $F$ and is derived from a given scenario of air temperature ($T$):

$$F = F_{ref} + \frac{F_{th} - F_{ref}}{T_{th} - T_{ref}}(T - T_{ref}) \tag{9}$$

where $F_{ref}$ and $T_{ref}$ are reference values. The detailed derivation of this model can be found in Stommel (1961) and Cessi (1994).

We also used the 4[th]-order Runge-Kutta method for numerical calculation. Timestep $dt(year)$ is 0.1 year and the simulation period is 1000 years. Whether the system tips or not was determined by the criteria $y > 0.99$ at the last step of the simulation. The details of the model setting are found in Table S4 and Table S5 in the Supporting information.

### 2.4 Experimental setting

As explained in Section 2.1, the 'true model' is a model which knows all the parameters and initial conditions without any uncertainties while the "imperfect model" has uncertainties. We prepared the "imperfect model" as a model assuming two of model parameters are uncertain. The two parameters are sensitive to whether the system will tip or not and they are chosen not to be easily estimated from just a small number of observations. At the initial time, two of model parameters were randomly sampled from a bounded uniform distribution (see Tables S3 and S5). The ensemble size of PF was set to 1000, so that we generated the 1000 combination of the model parameters at the initial time. By sequentially assimilating observations, these parameter samples as well as state variables were updated (see Equation (3)). We assumed no uncertainty in initial state variables since they do not significantly affect the long-term trajectory of our climate tipping models.

The forcing term of the two climate tipping models depends on temperature scenarios. Huntingford et al. (2017) suggested the simple temperature scenarios which approximates the temperature changes estimated by ESMs. We further simplify Hungtingford et al. (2017):

$$T(t) = \begin{cases} T_{st} + rt, & t < t_1 \\ T_{th} + dT_{ex}, & t_1 \leq t < t_2 \\ T_{th} + dT_{ex} - s(t - t_2), & t_2 \leq t < t_3 \\ T_e, & t_3 \leq t \end{cases} \quad (10)$$

where $r = \frac{T_{th} - T_{st}}{t_u}$, $t_1 = \frac{T_{th} + dT_{ex} - T_{st}}{r}$, $t_2 = t_1 + dt_{ex} - \frac{dT_{ex}}{s} - \frac{dT_{ex}}{r}$, $t_3 = t_2 + \frac{T_{th} + dT_{ex} - T_e}{s}$

The temperature starts from $T_{st}$(°C), and increases up to $dT_{ex}$(°C) over the temperature threshold $T_{th}$(°C). Then the temperature starts to decrease until it reaches $T_e$(°C).

We also make it possible to change internal variability in the temperature profile. The internal variability is expressed by giving random gaussian noises in each timestep.

$$T_{var}(t) = T(t) + \sigma_{sig} W_t \quad (11)$$

where $\sigma_{sig}$ is a parameter which controls the amplitude of internal variability and $W(t)$ is the random process and at each step it follows normal distribution N(0, dt). The temperature scenarios and the corresponding output of the models are shown in Figure S1 and Figure S2.

It was assumed that $v$ in the simplified TRIFFID model and $Q$ in AMOC two box model were observed. Observation data were generated by adding Gaussian white noises to $v$ and $Q$ of the nature runs. The mean of the Gaussian white noise is set to 0, so that observation is unbiased. By changing the variance of the Gaussian white noise, we changed the observation error and assessed the impact of observation error on the predictability of climate tippings. We also changed observation frequency and the time when last observation is available to check their impact on predictability.

The practical predictability was evaluated by counting the number of ensemble members which accurately identify climate tipping (non-tipping), if the nature run shows tipping (non-tipping). Then, the ratio of the number of these accurate members to the total ensemble size (i.e., 1000) is calculated as the evaluation metrics. If the ratio is 1, all the ensemble members accurately show tipping or non-tipping, on the other hand, if the ratio is 0, the model prediction of climate tipping is completely wrong. This accuracy is influenced by the randomness of the observation noise, so that we performed 100 trials. We used the mean of the accuracy metrics of these 100 numerical experiments.

We divided the experiments into two parts (see Table S1). In the experiment 1, we examined the necessary accuracy and frequency of observations to improve model predictions by data assimilation as a function of the magnitude of the internal variability of the system. The last observation is fixed to two years before the temperature without internal variability reaches the threshold where the system loses its stability. We conducted experiments with 5 different internal variabilities ($\sigma_{sig}$ in Equation (11)), 5 different observation errors, and 6 patterns of observation frequency. Totally we performed 5*5*6 = 150 numerical experiments.

In the Experiment 2, we examined the lead time of the prediction of the climate tippings as a function of observation accuracy and internal variability of the system. We conducted experiments for 5 different internal variabilities, 3 (4) different observation errors in the simplified TRIFFID model (in the AMOC two box model), and 4 observation periods. Observation period is defined as the duration from the first to the last observations. While the first observation timing is fixed, the last observation timing was changed. As observation period gets shorter, it is necessary to predict the existence of climate tipping longer before the temperature steps across the threshold of tipping. The frequency of observation is fixed to once in 2 years. The detailed setting of parameters in these experiments are shown in Table S1-S5 in Supporting information.

## 3. Results

We demonstrate the result of OSSE showing several numerical experiments in the experiment 1. Figure 1 shows the timeseries of $v$ in the simplified TRIFFID model and the estimated parameters at the end of simulation. When we have no observation, the estimation of $v$ has extremely large uncertainty due to the large initial uncertainty in the model parameters, and the existence of tipping could not be accurately estimated (Figures 1a and 1d). Although observation can constrain model parameters, noisy observation is not useful to accurately predict the existence of climate tipping (Figures 1b and 1e). When observation is sufficiently accurate, we can accurately predict the synthetic true climate tipping (Figure 1c) and the synthetic true model parameters (Figure 1f).

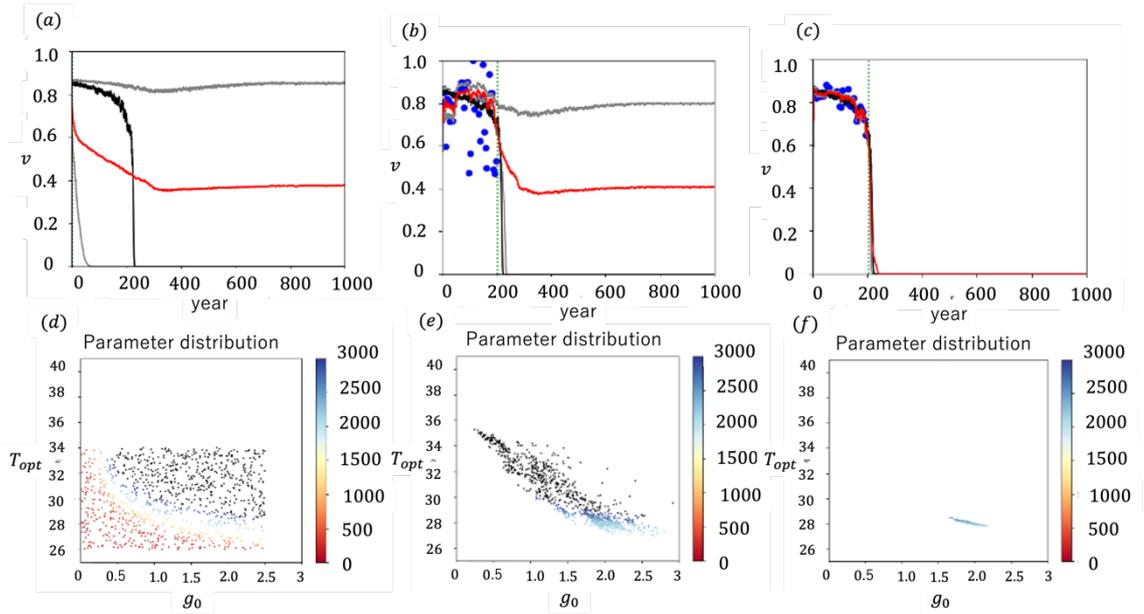

**Figure1**: (a,b,c) The timeseries of Amazon rainforest ratio $v$. Vertical and horizontal axes are forest ratio and time, respectively. Black and red lines are nature run (the synthetic true rainforest ratio) and ensemble mean of model prediction, respectively. Two grey lines are a line connecting the first quartiles at each time and that connecting the third quantile at each time. Blue dots are observation. (d,e,f) The distribution of the model parameters at the end of the numerical simulations. Vertical and horizontal axes are $T_{opt}$ and $g_0$, respectively (see Section 2.3.1). Color of each dot shows the timestep of tipping for each ensemble member. The timestep can be interpreted as year by multiplying dt=0.1. Black dots are the parameters which do not show tipping till the simulation ended. (a,d) Ensemble simulation with no data assimilation. (b,e) the case where observations with 20%-error are assimilated. (c,f) the case where observations with 2.5%-error are assimilated.

Figure 2 summarizes the results of the experiment 1. The skill to predict tipping can be improved with smaller observation errors. Even under the same observation error, the skill to predict climate tipping is improved by increased internal variability of temperature. For example, in the case of the simplified TRIFFID model, If the internal variability amplitude $\sigma_{sig}$ is smaller than 2 [K], observations with 5% errors are not beneficial to improve the model prediction of climate tipping (Figures 2a, 2b, and 2c). On the other hand, if the internal variability amplitude $\sigma_{sig}$ is larger than 2[K], the observation data can greatly improve the skill of model prediction (Figures 2d and 2e). This tendency can also be found in the AMOC two box model (Figures 2f-2j). When the internal variability is large, even less accurate observation is effective to improve the model prediction of climate tipping. Compared to the observation error, the observation frequency does not greatly affect the skill to predict climate tipping. Observations in every 20 years are frequent enough to predict climate tipping. It should be noted that it is easier to improve the predictability of the non-existence of climate tipping by data assimilation (Figure S3). Since data assimilation can improve the model prediction in the cases of both non-tipping and tipping scenarios, our data assimilation method is not biased to inform that climate tipping will happen.

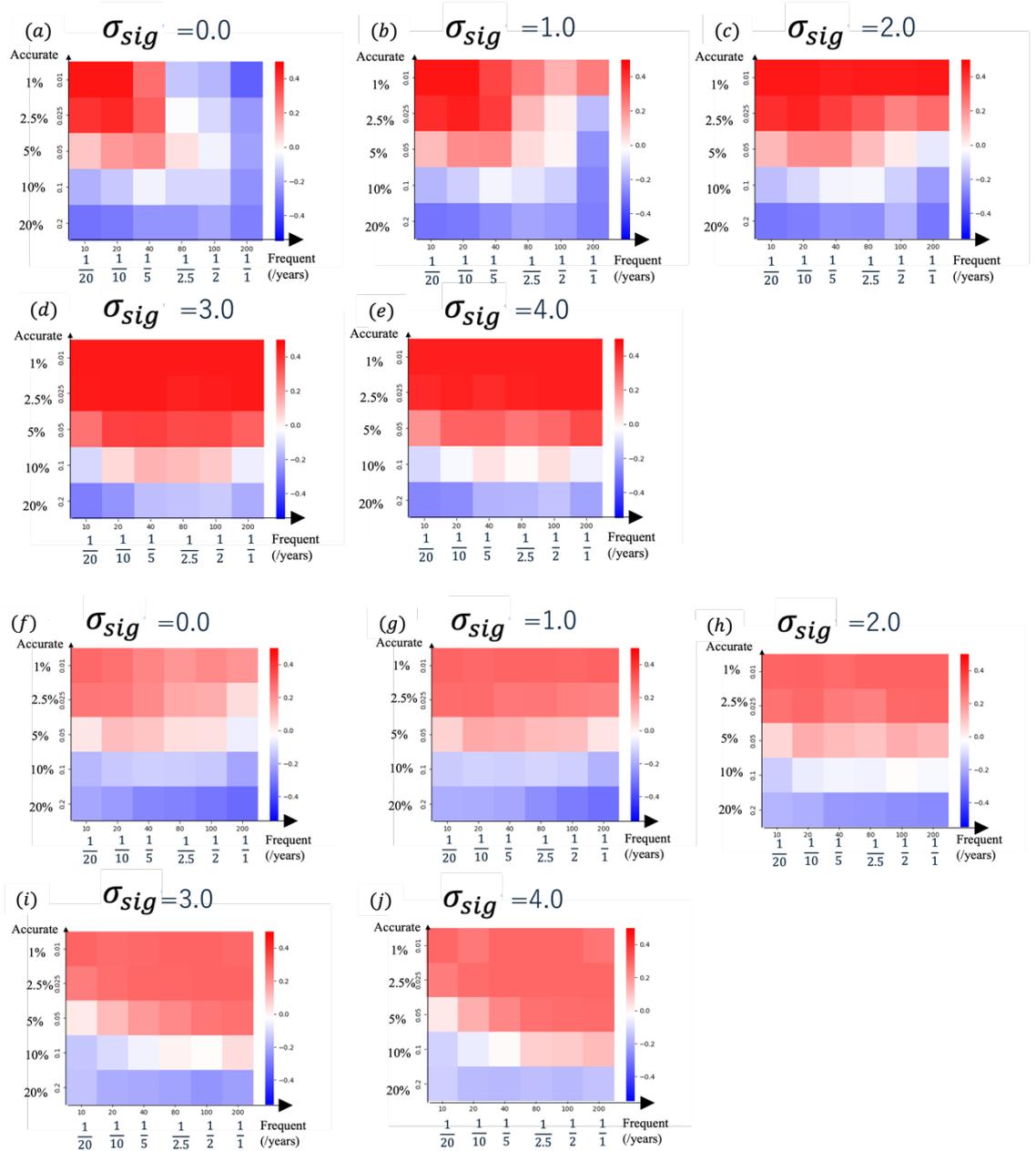

Figure 2 : The difference of the ratio of the number of accurately predicted ensemble members to the total ensemble size between data assimilation and no data assimilation experiments under climate tipping scenario. Red color indicates that the prediction accuracy is improved by the observation data. Blue color indicates that the prediction accuracy is degraded by the observation data. (a-e) The results of the simplified TRIFFID model. (f-j) The results of the AMOC two box model. Each model's result includes 5 internal variability settings. Vertical and horizontal axes are observation error and frequency, respectively.

Figure 3 summarizes the result of the experiment 2. Toward appropriate climate policy making, it is necessary to predict the existence of climate tipping under a specific climate scenario with the long lead time. However, it is difficult to predict climate tipping with long lead time especially when observation error is large and internal variability is small. As we found in Figure 2, the system's internal variability and observation error substantially affect the skill of data assimilation to predict climate tipping. When lead time is long, we can have the access to observation only in the short period when the system is in the stabilized state. In this case, it is difficult to learn model parameters which induce instability of the system and climate tipping. Large internal variability and accurate observation enable to learn model parameter even in this case of long lead time.

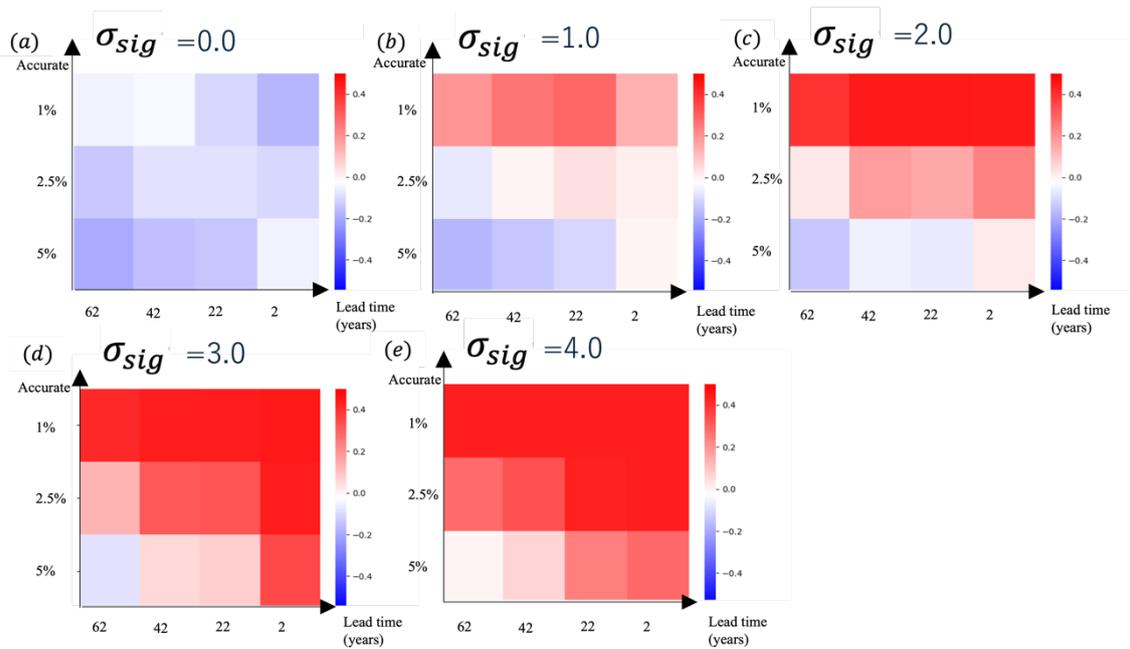

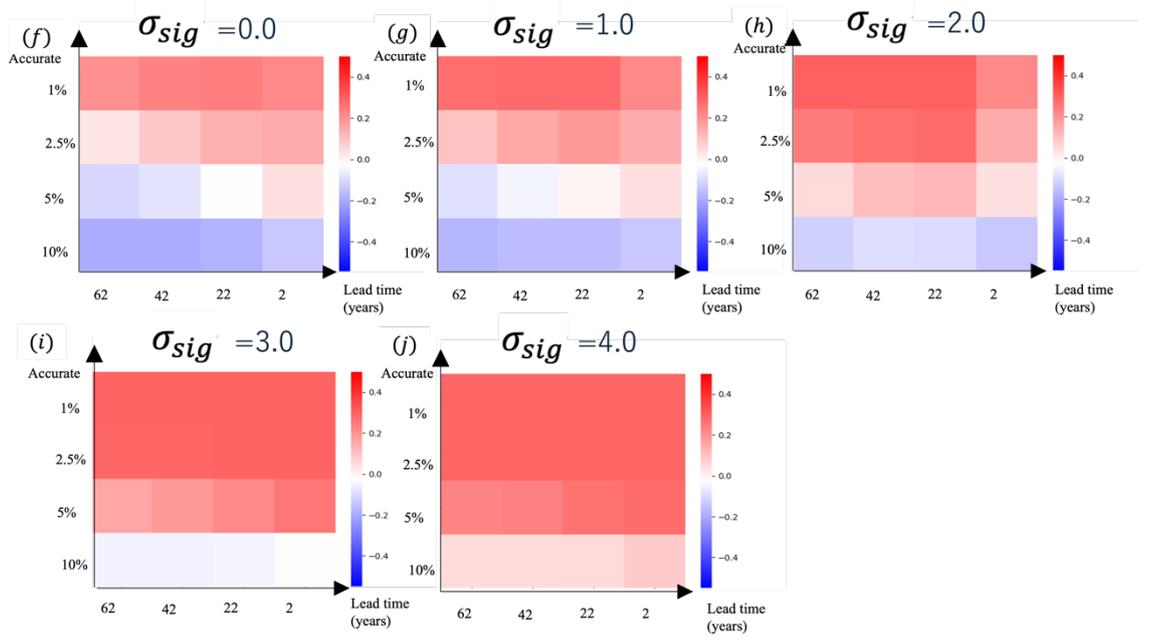

**Figure 3**: The difference of the ratio of the number of accurately predicted ensemble members to the total ensemble size between data assimilation and no data assimilation experiments. Red color indicates that the prediction accuracy is improved by the observation data. Blue color indicates that the prediction accuracy is degraded by the observation data. (a-e) The results of the simplified TRIFFID model. (f-j) The results of the AMOC two box model. Each model result includes 5 internal variability setting results. Vertical and horizontal axis are observation accuracy and lead-time, respectively. The lead-time is measured as the period from the last observation timestep to the point when the system loses stability.

The experiments 1 and 2 imply that the magnitude of internal variability of the system relative to observation error is important. We define S/N ratio to sort out all the results above. The S/N ratio is defined as:

$$S/N = \frac{\max_{t \in [t_0, t_{end}]} \overline{\sigma_{sig}}(t)}{\sigma_{obs}}, \qquad (12)$$

where $\overline{\sigma_{sig}}(t)$ is the amplitude of internal variability of which is the standard deviation of linearly-detrended true state ($v$ and $Q$) timeseries from t-30 (year) to t (year), $\sigma_{obs}$ is observation noise, and $t_0$ is the 50th step of the simulation. We used the 50th timestep instead of the first step, because the spin-up period should be excluded to avoid the effect of inaccurate specification of the initial state. $t_{end}$ is the last step of the period when observation is available. Figure 4 clearly shows that the larger

S/N ratio provides the accurate model prediction of climate tipping and if the S/N ratio is larger than 1, the model accurately predicts the climate tipping. Even if climate tipping and the critical phase change of the system are not directly observed, the existence and timing of climate tipping can be predicted if the relevant internal variability of the system can be accurately observed. Figure 4 provides the prototype to investigate the necessary S/N ratio of observation to effectively constrain the model parameters of the process-based models to accurately assess the climate tipping.

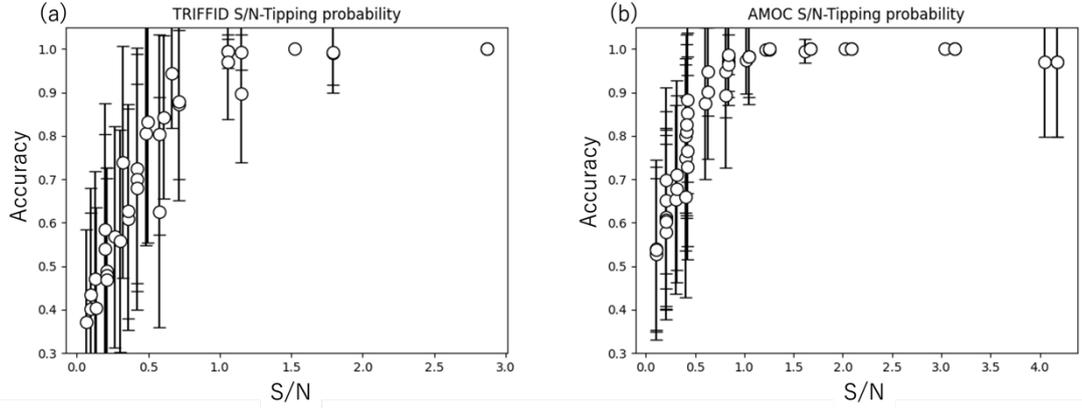

**Figure 4**: The relationship between the prediction accuracy of climate tipping and the S/N ratio in (a) the simplified TRIFFID model and (b) the AMOC two box model. Here we plotted only tipping -scenario results in all numerical experiments. The error bar shows the variance of prediction accuracy obtained from 100-time experiments.

4. **Discussion & conclusions**

We explore the practical predictability of climate tipping focusing on the internal variability of the system and the observation accuracy. Inspired by OSSE in numerical weather prediction, we propose the framework to evaluate the practical predictability of climate tipping. We find that the internal variability of the systems and the observation error determine the predictability of climate tipping. Even if direct observation of climate tipping is absent, the observation which is accurate enough to capture the relevant internal variability is useful to improve the skill of process-based models to predict the existence and timing of climate tipping.

Our analysis of the relationship between S/N ratio and the accuracy of predicting climate tipping (Figure 4) can provide the criteria of beneficial observations to improve the skill to predict climate tipping. Here we demonstrate how to use our finding to evaluate the existing observation in terms of the practical predictability of climate tipping. The S/N ratio defined in Equation (12) is the ratio of observed internal variability standard deviation to the observation error. Note that observed internal variability consists of true internal variability and observation error, so that the S/N ratio is overestimated when both internal variability and observation error are obtained from a single observation. Atkinson et al. (2011) investigated satellite observation of Normalized difference vegetation index (NDVI) and Enhanced vegetation index (EVI) in amazon from 2000 to 2010. In their assessment, the interannual variability of EVI is around 0.03 and observation root mean squared error is around 0.01. S/N ratio is derived just by taking the ratio of the internal variability and observation error, 0.03/0.01=3. This is larger than the necessary S/N ratio to predict climate tipping shown in Figure 4. McCarthy et al. (2015) mentioned that the Mooring Array observation from April in 2004 to October in 2008 shows the standard deviation of the flow rate from the mean is found to be 0.9 Sv. They reported that the error in this observation network was about 1.5 Sv. In this case, S/N ratio is 0.9/1.5=0.6, which is smaller than the critical S/N ratio (i.e., 1.0) shown in Figure 4. The existing AMOC observation is not accurate enough to make the model accurately predict its tipping.

We used the simple and stylized models. Although these simple models have recently been used to qualitatively evaluate the behaviors of climate tipping (Ritchie et al. (2021), Lohmann et al. (2021), Wunderling et al. (2021)), they overly simplified the complex processes of climate dynamics. It is more beneficial to perform our proposed OSSE to measure practical predictability of climate tipping with more complicated models considering more complex model imperfectness. This study is the first step to evaluate the value of observation to improve process-based models' prediction of climate tipping. Future work should focus on practical predictability of climate tipping using realistic ESMs.

**Acknowledgements**

This work was supported partly by JST Moonshot R&D program (JMPJMS2281) and JSPS KAKENHI grant (21H01430).

**Open Research**

This study was a theoretical work, so that no dataset was used. The source code is available at: https://zenodo.org/records/12079800

**Text S1.**

Particle filter algorism

The step-by-step description of our particle filter algorithm is written below.

(1) N samples of parameter $\theta(t)$ are initially taken from appropriate uniform distribution.

(2) Derive prior distribution of variable $x(t)$ and parameter $\theta(t)$

($\Pr[X = x(t), \Theta = \theta(t)|Y = y(1:t-1)]$) by time evolution according to the system model: $x(t) = f(x(t-1), \theta(t-1)) + \epsilon(t)$.

(3) Derive likelihood to the observation data on each sample. (Likelihood of sample i is $w_i$ and cumulative likelihood is $W_i \equiv \frac{\sum_{k=1}^{i} w_k}{\sum_{k=1}^{N} w_k}$)

(4) Generate N random samples $\{r_j\}$ (j = 1,2, ..., N) from U [0,1] and if $W_i \leq r_j < W_{i+1}$ (i = 0, 1, ..., N − 1), gain sample i+1.

(5) Derive parameter distribution covariance matrix as $\Sigma$, and add noise on the samples which follows normal distribution whose mean is 0 and covariance matrix is $a^2\Sigma$. (a = 0.1 in this research). Go back to (2).

Steps (1) and (2) are the processes to approximate prior distribution, $p(x(t), \theta(t)|y(1:t-1))$. Steps (3) and (4) are the processes to update the distribution and resample the posterior distribution. Step (5) is the inflation process which prevents the degeneracy in which all but one ensemble members have nearly zero weights.

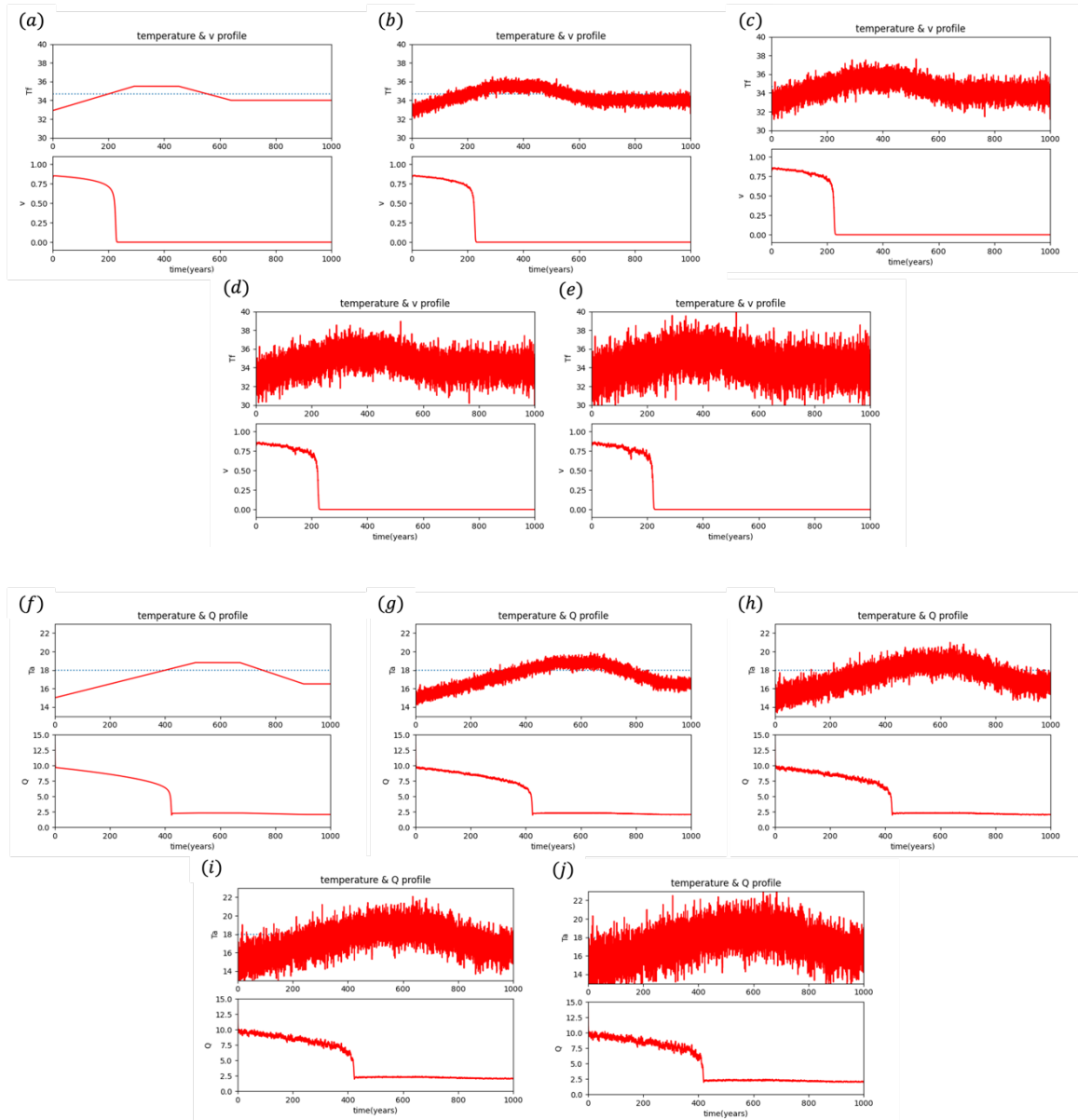

**Figure S1.** Timeseries of temperature and estimated states of the models used in this experiment. (a-e) The case with tipping scenarios of the simplified TRIFFID model. (f-j) The case with tipping scenarios of the AMOC two box model. The upper figure shows the timeseries of temperature which is the external forcing and the lower figure shows the timeseries of model's state variables. The states of the two models are $v$ in the simplified TRIFFID model and $Q$ in the AMOC two box model. The size of internal variability increases from (a) to (e) and from (f) to (j) ($\sigma_{sig} = [0, 1, 2, 3, 4]$).

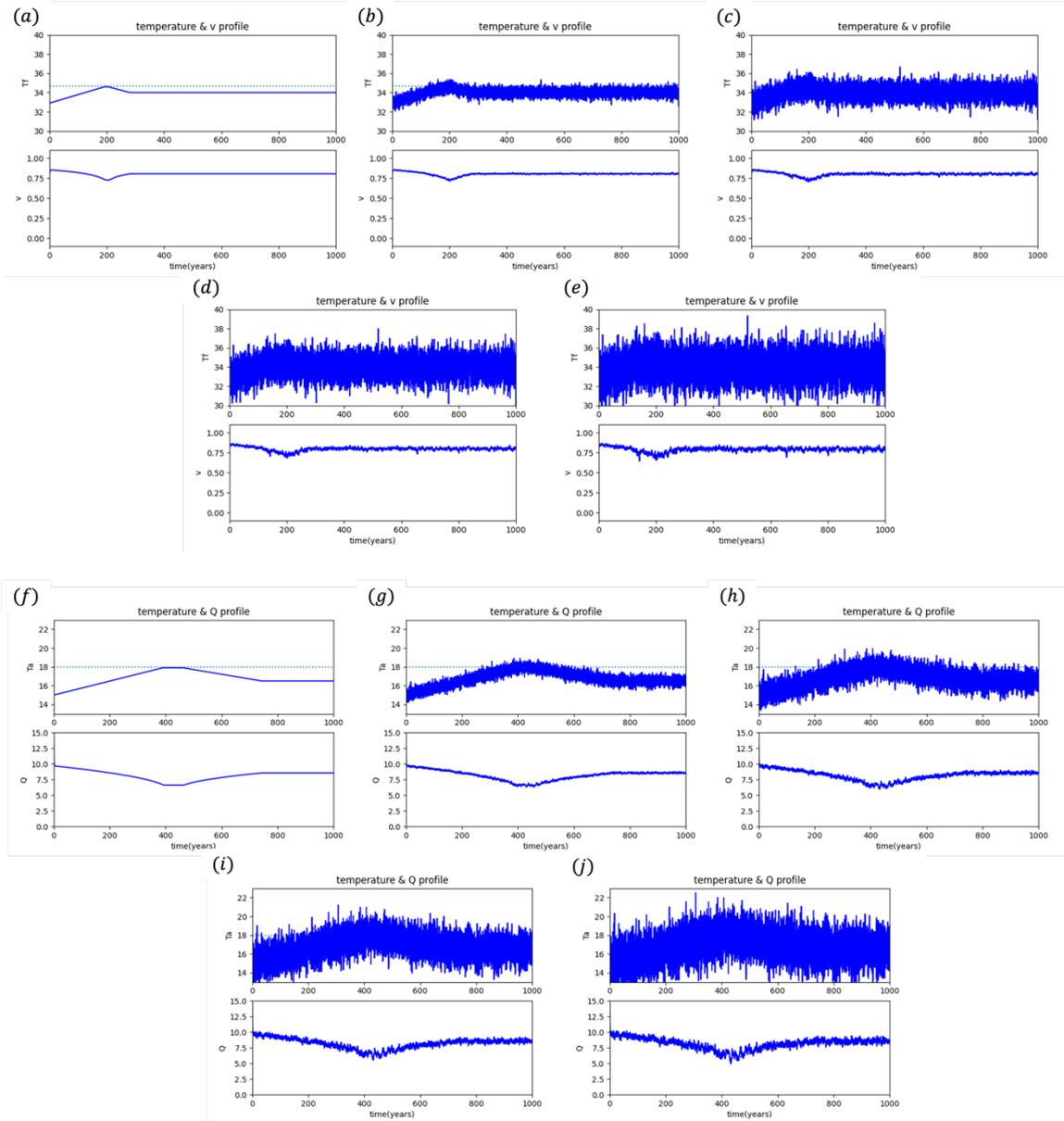

**Figure S2** Timeseries of temperature and estimated states of the models used in this experiment. (a-e) The case with non-tipping scenarios of the simplified TRIFFID model. (f-j) The case with non-tipping scenarios of the AMOC two box model. The upper figure shows the timeseries of temperature and the lower figure shows the timeseries of model's state variables. The states of the two models are $v$ in the simplified TRIFFID model and $Q$ in the AMOC two box model. The size of internal variability increases from (a) to (e) and from (f) to (j) ($\sigma_{sig} = [0, 1, 2, 3, 4]$).

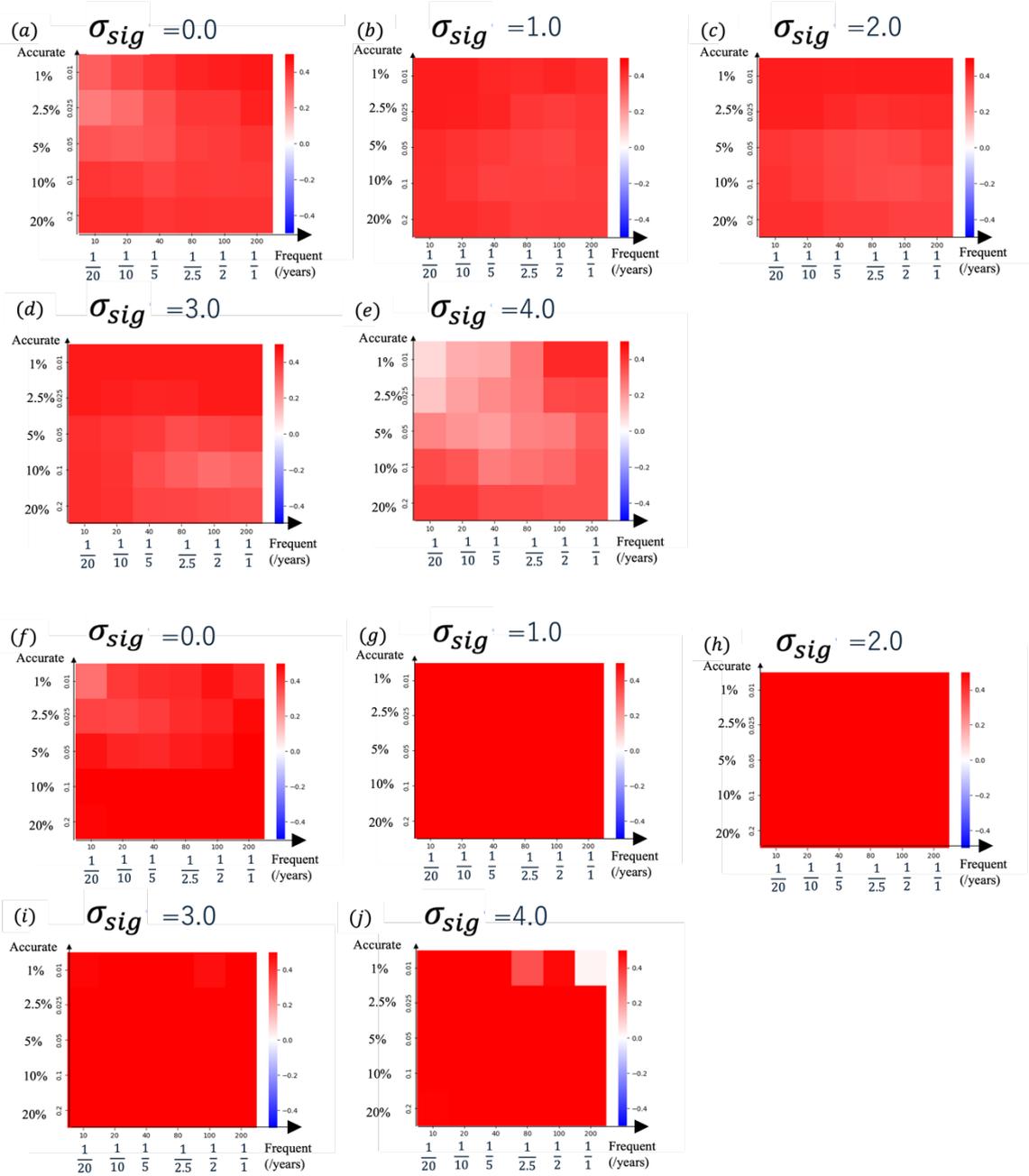

**Figure S3** The difference of the ratio of the number of accurately predicted ensemble members to the total ensemble size between data assimilation and no data assimilation experiments. Red color indicates that the prediction accuracy is improved by the observation data. Blue color indicates that the prediction accuracy is degraded by the observation data. (a-e) and (f-j) are the results of the simplified TRIFFID model and the AMOC two box model with tipping scenario, respectively. Each model result includes 5 internal variability settings. Vertical and horizontal axes are observation error and frequency, respectively.

**Supporting information**

**Table S1.** Experiment setting.

| Experiment | Model | Observation | Internal variability |
|---|---|---|---|
| Experiment1 | TRIFFID | Error [0.01, 0.025, 0.05, 0.1, 0.2] Frequency [$1/1, 1/2, 1/{2.5}, 1/5, 1/{10}, 1/{20}$ (/years)] Last observation step 2 (years before losing its stability) | $\sigma_{sig}$ = [0,1,2,3,4] |
| | AMOC | Error [0.1, 0.25, 0.5, 1.0, 2.0 (Sv)] Frequency [$1/1, 1/2, 1/{2.5}, 1/5, 1/{10}, 1/{20}$ (/years)] Last observation step 2 (years before losing its stability) | $\sigma_{sig}$ = [0,1,2,3,4] |
| Experiment2 | TRIFFID | Error [0.01, 0.025, 0.05] Frequency $1/2$(/years) Last observation step [2, 22, 42, 62(years before losing its stability)] | $\sigma_{sig}$ = [0,1,2,3,4] |
| | AMOC | Error [0.01, 0.025, 0.05, 0.1 (Sv)] Frequency $1/2$(/years) Last observation step [2, 22, 42, 62(years before losing its stability)] | $\sigma_{sig}$ = [0,1,2,3,4] |

**Table S2.** Simplified TRIFFID scenario setting

| Parameter | Meaning | Tipping scenario | Not tipping scenario |
|---|---|---|---|
| $T_{st}$ | Temperature at the start point of simulation (°C) | 32.9 | 32.9 |
| $T_{th}$ | The temperature threshold of the system losing its stability (°C) | 34.7 | 34.7 |
| $T_e$ | Temperature after the simulation (°C) | 34 | 34 |
| $s$ | The speed of temperature decrease after reaching maximum temperature $T_{th} + dT_{ex}$ (°C/year) | 0.008 | 0.008 |
| $t_u$ | The time of the system losing its stability (year) | 202 | 202 |
| $dT_{ex}$ | The difference of maximum temperature from $T_{th}$ (°C) | 0.8 | 0.01（without internal variability）<br>-0.1（with internal variability） |
| $dt_{ex}$ | The period of the temperature over $T_{th}$ (year) | 350 | 40（without internal variability）<br>-10（with internal variability） |

**Table S3.** Simplified TRIFFID parameter setting

| Parameter | Meaning | True model | Imperfect model |
|---|---|---|---|
| $T_{opt}$ | The temperature when the growth rate is at maximum (°C) | 28 | Uniform distribution [26,34] |
| $g_0$ | The maximum growth rate | 2 | Uniform distribution [0,2.5] |
| $\alpha$ | The reference temperature of the rainforest ratio's effects (°C) | 5 | 5 |
| $\beta$ | The reference temperature of the growth rate decrease (°C) | 10 | 10 |
| $\gamma$ | Disturbance rate | 0.2 | 0.2 |

**Table S4.** AMOC two box model scenario setting

| Parameter | Meaning | Tipping scenario | Not-tipping scenario |
|---|---|---|---|
| $T_{st}$ | Temperature at the start point of simulation (°C) | 15 | 15 |
| $T_{th}$ | The temperature threshold of the system losing its stability (°C) | 18 | 18 |
| $T_e$ | Temperature after the simulation (°C) | 16.5 | 16.5 |
| $T_{ref}$ | The reference temperature to convert temperature scenario to fresh water flux (°C) | 16 | 16 |
| $s$ | The speed of temperature decrease after reaching maximum temperature $T_{th} + dT_{ex}$(°C/year) | 0.010 | 0.005 |
| $t_u$ | The time of the system losing its stability(year) | 402 | 402 |
| $dT_{ex}$ | The difference of maximum temperature from $T_{th}$ (°C) | 0.8 | 0.01 (without internal variability) <br> -0.1 (with internal variability) |
| $dt_{ex}$ | The period of the temperature over $T_{th}$(year) | 350 | 40 (without internal variability) <br> -10 (with internal variability) |
| $F_{th}$ | The freshwater flux threshold of the system losing its stability (°C) | 1.296 | 1.296 |
| $F_{ref}$ | The reference freshwater flux which is corresponding to the reference temperature$T_{ref}$(Sv/year) | 1.1 | 1.1 |

**Table S5.** AMOC two box model parameter setting

| Parameter | Meaning | True model | Imperfect model |
|---|---|---|---|
| $\eta$ | Scaling parameter | $3.17 \times 10^{-5}$ | Uniform distribution $[2.0 \times 10^{-5}, 5.0 \times 10^{-5}]$ |
| $\mu$ | The ratio of timescale of advection to that of diffusion | $\sqrt{6.2}$ | Uniform distribution $[0.5, 3.5]$ |
| $V$ | The volume of the ocean ($km^3$) | $300 \times 4.5 \times 8250$ | $300 \times 4.5 \times 8250$ |
| $t_d$ | The timescale of diffusion(year) | 180 | 180 |